# Towards an understanding of particle-scale flaws and microstructure evolution in cold-spray via accumulation of single particle impacts

*Alain Reiser[1,2], Christopher A. Schuh[1,3]*

[1]Department of Materials Science and Engineering, Massachusetts Institute of Technology, Cambridge, MA 02139, USA

[2]Department of Materials Science and Engineering, KTH Royal Institute of Technology, 114 28 Stockholm, Sweden

[3]Department of Materials Science and Engineering, Northwestern University, Evanston IL 60208, USA

E-mail: schuh@northwestern.edu

**Cold spray coatings are the sum of countless individual bonding events between single particles impacting on top of one another at high velocities. Thus, the collective behavior of microparticles must be considered to elucidate the origins of coating flaws at the scale of the particles and larger, or the dynamic evolution of the overall coating microstructure. Laser-induced particle impact testing (LIPIT) has been extensively used to study single-particle impacts, and in this work is adapted to study the accumulation of numerous particles with knowledge of each individual particle's impact parameters (particle size, velocity). The method reproducibly deposits stacks of gold particles (>20 particles) with different characteristic spectra of impact velocity. The observation of impact-induced erosion lets us define a critical velocity for material-build-up that is higher than that for single-particle bonding. The quantitative single-particle data are analyzed in a correlative manner to the structure and flaws in the resulting stacks, providing some first statistical connections between, e.g., strain and recrystallization, or aberrant particle characteristics and defects. The results highlight opportunities for the study of many-particle phenomena in microparticle impact—from interaction of particles in cold spray to multi-step erosion processes—with a quantitative view of the behavior of single particles.**

## 1. Introduction

Cold spray is an advanced manufacturing technology for coating, repair, and increasingly, the additive manufacturing (AM) of engineering metals[1,2]. In cold spray, thick coatings are formed via acceleration of micron-sized particles to supersonic velocities, and their subsequent kinetic bonding onto a substrate upon impact. Fundamentally, a coating is the result of myriad bonding events of individual metal microparticles. Bonding of single particles is generally associated with a critical particle impact velocity—sufficient kinetic energy must be available to perform the plastic work necessary for solid-state bonding. Consequently, extensive studies of single-particle impacts have been conducted to better understand the fundamental aspects of these impacts.

Traditionally, single-particle impacts are studied with a wipe test. Here, a cold-spray plume with a known average particle size and velocity is rapidly translated across the substrate so as to deposit less than a monolayer of coating with isolated particles exposed on the substrate. This approach is straightforward and simple but lacks truly quantitative character, as the individual particles' impact parameters (original size, inbound velocity, and temperature) are unknown. A more recent technique is laser-induced particle impact testing (LIPIT)[3,4], where single particles are accelerated by laser-ablation, mimicking the cold spray process in a tabletop optical setup. The synchronization of particle impact with high-speed videography enables the observation of the impact event and the measurement of quantitative data such as inbound and outbound velocities, and thus the kinetic energy dissipated in an impact event. The combination of LIPIT with post-mortem analysis by optical and electron-optical microscopy allows the study of particle deformation, fragmentation and bonding mechanisms. As a result, LIPIT has provided critical velocities for bonding of a wide range of metal particles[3,5,6], elucidated the influence of oxide layers[7,8], directly linked bonding to hydrodynamic jetting of material at the particle rims[9–11], mapped the deformation modes of particles[12] as well as the onset of erosion and melting[13,14], and revealed novel mechanisms for dynamic and static recrystallization upon or after impact [15,16]. LIPIT has thus proven to be an extremely versatile approach for the study of a wide array of phenomena related to cold spray, as it provides a quantitative, single-particle view of the impacts that underlie it.

However, such single particle experiments do not speak to the inherently multi-particle phenomena known to influence cold-spray coating structure and properties. These include the tamping (densification) and peening (cold-working) of previously-deposited portions of the coating by subsequent particle impacts[2,17], the inclusion of pores between particles, or the fracture and erosion of coating material by particle strikes[17–20]. These collective phenomena are typically studied only at a coating-level, but without access to the particle-level details that have proven so powerful in single-impact LIPIT experiments.

There is therefore a clear need for a single-particle view of the many-particle process that forms a cold spray coating, and our purpose in this paper is to provide a first step towards that goal. We employ LIPIT to study the collective behavior of gold microparticles forming many-particle deposits (up to 24 particles) through a coordinated sequence of individual shots with full LIPIT-accuracy on each shot. Such data are shown to offer better understanding of the missing links between inter-particle defects, coating microstructure, and individual particle behavior.

## 2. Experimental

### 2.1. Laser-induced particle impact testing (LIPIT)

Detailed descriptions of the LIPIT setup have been published elsewhere[21]. Launch pads—the substrates that launch microparticles at high velocities—were composed of a 210 μm thick glass substrate (Corning No. 2 microscope cover slip, 25 mm diameter) coated with a 90 nm thick chromium and 75 μm thick polyurea layer. Polyurea precursors (Modified MDI Isocyanate curative, RCS Rocket Motor Components, and GCLink P-650, Chem Coast Incorporated) were mixed in a planetary centrifugal mixer (2 min, ARE-310, Thinky) in a mass ratio 1:2.4 (MDI:P-650) and then spin-coated (G3P-8, Cookson Electronic Equipment) at 750 rpm for 5 min. The PU films were cured at 85 °C for 24 hrs under vacuum. Gold microparticles (99.9%, Alfa Aesar, 200 mesh) were spread on the polymer layer using lens cleaning papers and a drop of ethanol. For impact-bonding experiments, gold targets were placed at a distance of 150-200 μm from the launch pad. Two kinds of targets were used, Si wafers coated with ca. 100-nm thick, sputter-deposited gold films, and Si wafers electroplated with ca. 3-μm-thick gold films (TSG-250, *Transene Company*).

Particle stacks, or "deposits", were formed by sequentially impacting and bonding particles without translation of the target in-between shots. To launch selected particles (9–20 μm in diameter) towards the target, an intense laser pulse (pulsed Nd:YAG, pulse width 10 ns, λ=532 nm) was focused onto the launch pad (30-mm focal length lens with a minimal focal spot size of 5 μm). Upon ablation of the metal film, rapid expansion of the generated plasma bulges the polymer film and in turn accelerates and launches one selected microparticle towards the target. The particle speed was controlled by adjusting the laser energy between typically 0.4 and 0.7 mJ. The particle size was measured before launch, and after each successful launch a new region of the launchpad was used to isolate and then launch the next particle.

To reproducibly overlap impacts and successfully build up many-particle stacks, reliable alignment of particles in the center of the ablation laser beam is crucial. In previous versions of the LIPIT system, a single imaging path was used for this task. For this study we used a second imaging path perpendicular to the first to render particle alignment more precise[22]. The growth and fracture of deposits was also observed with the same imaging path used for particle alignment. The launch and impact of individual particles was imaged with a high-speed camera (SIMX16, *Specialised Imaging*), with a typical frame and interframe time of 5 and 45–95 ns, respectively. Particle velocities were measured from these image sequences. An error in velocity measurement of ±5% was estimated based on multiple re-measurements of a range of image sequences at the relevant interframe times. This fractional error is displayed by the error bars. A 10 μs laser pulse (λ=640 nm) was used for illumination of 16 frames. The magnification of each camera was calibrated with resolution targets before experiments.

## 2.2.   Analysis
The dimensions of particle deposits were measured with a laser scanning confocal microscope (VK-X250, *Keyence*). SEM was performed with a Gemini 450 instrument (*Zeiss*). Cross-sections of deposits were prepared in a dual-beam FIB (Helios 660, *FEI*) with a typical final milling current of 9.3 nA. Areas and lengths in pore and microstructure maps were measured in Adobe Illustrator using a script by Kate Baldwin (https://www.k8baldwin.com/). Linear regression analysis was performed with *IBM SPSS Statistics*.

## 3.   Many-particle deposits with quantitative, single-particle impact data
In LIPIT, individual microparticles of known size are launched towards a target at up to 1000 m s⁻¹ while their impact velocity is monitored by high-speed videography (**Figure 1a**). The top image sequence in Figure 1b shows a gold particle, 14 μm in diameter, impacting a gold target at 358 m s⁻¹. The absence of a rebound indicates kinetic bonding of the particle to the target. Adhesion is expected at this impact velocity, as the critical velocity for bonding of gold particles has previously been measured to be on the order of 250 m s⁻¹ (for particles of diameter 16 ± 4 μm)[5]. In traditional single-impact studies by LIPIT, the target would now be moved to present a pristine impact site for the next test. In our case, however, the target is not translated in-between single shots. Thus, the next particle hits and bonds close to or on top of the previously adhered particle. A 3D deposit is formed upon the continued repetition of this process (Figure 1c). As the interparticle time is on the order of minutes, there is no accumulation of heat produced by plasticity; all particle launches occur under static, isothermal initial conditions.

The formation of the particle stack is observed with two levels of temporal resolution. On one hand, the impact of every particle is resolved with high-frame-rate imaging. The bottom image sequence in Figure 1b depicts the impact-bonding of the 24th gold particle onto a previously deposited stack of particles. Just as with single-particle impacts, such an image sequence reveals inbound and rebound velocities of particles (if there is rebound). On the other hand, post-impact photography shows the growth of the entire deposit (Figure 1d). Importantly, the latter resolves fracture of the deposits (fracture typically occurs on a longer time scale than what the high-frame-rate imaging captures).

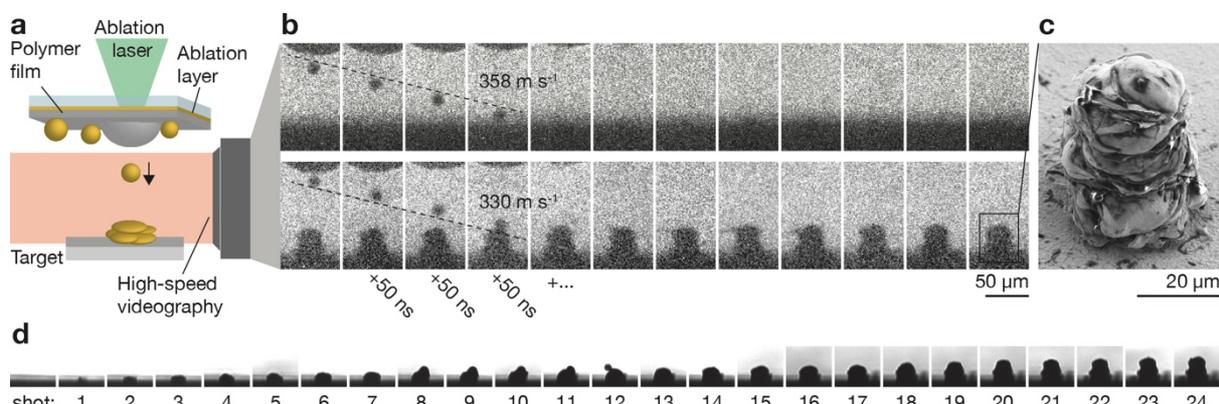

**Figure 1. Many-particle deposition by laser-induced particle impact testing (LIPIT).** **a)** In LIPIT, individual microparticles are launched towards a target by laser ablation of a sacrificial metal layer and subsequent, rapid expansion of a polymer film. Sequential impact at the same target location enables the deposition of coatings or stacks of many particles. **b)** Synchronized, high-speed videography is used for the direct observation of the particles' trajectory and impact and the measurement of particle velocities. The sequences capture the impact-induced adhesion of single microparticles on a pristine target (top) and a stack of many particles (bottom). **c)** SE-SEM micrograph of a deposit of 24 gold microparticles. **d)** Post-impact optical micrographs for each of the 24 shots show the growing deposit.

Typical deposits studied here contain 20-30 particles (9–22 μm in diameter) and have an aspect ratio greater than one (**Figure 2**). The stack diameter of 30–50 μm is 2–4 times the original size of the particles (9–22 μm). The width of the deposit is caused by a combination of particle flattening upon impact and a lateral offset between particles. The offset, which can be as large as a particle diameter, is a result of the inherent experimental variability of launch trajectories in LIPIT. Conveniently, this natural variation simulates the range of particle offsets expected in actual cold-spray coatings.

In general, the quantitative view of every impact that builds a deposit is what crucially distinguishes the LIPIT approach from wipe tests or standard coating experiments. Based on single-particle data, the spectrum of particle sizes, velocities and dissipated kinetic energy of every deposit can be measured. **Figure 3** shows four successful particle stacks and the associated particle sizes and velocities (error bars: ±5%) of every particle contained therein. The average particle velocities ranged from 328±45 to 354±37 m s$^{-1}$, well above the critical velocity $v_{cr}$=253±7 m s$^{-1}$ for bonding of Au particles on a matching, flat, bulk substrate (reported in prior work for particles of size D=16 ± 4 μm)[5]. However, with LIPIT, we can go beyond the averaged view and assess local variability within the stacks. For example, we measure that the lowest particle velocity for bonding was 249 m s$^{-1}$ for a 20-μm particle; that one particle impacted at only 169 m s$^{-1}$ and probably rebounded (unfortunately, the data is inconclusive as the possible rebound event occurred after the acquisition window); and that some particles impacted at >400 m s$^{-1}$. Based on this particle-specific data, in principle, it should be possible to track the location of the particle within the coating and correlate possible microstructural features associated with it. We turn our attention to such correlative analysis in the following sections.

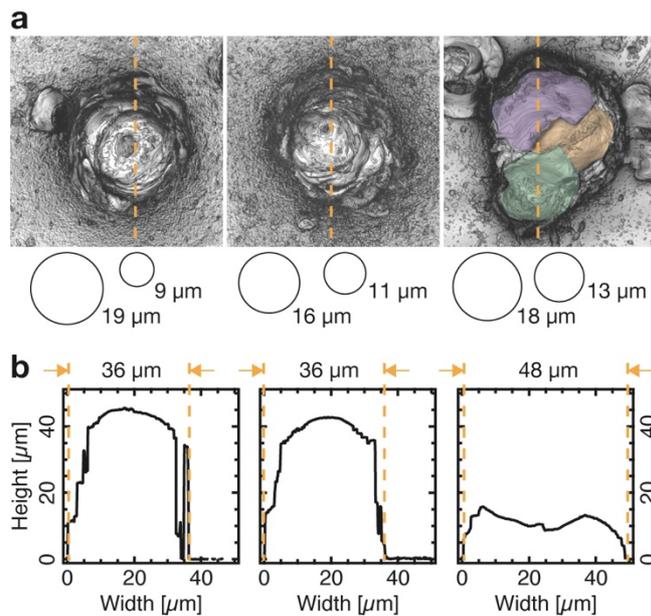

**Figure 2. Stacking of particles—lateral offsets.** Particles are laterally offset within the deposits due to inherent experimental variability of launch trajectories in LIPIT. **a)** Top-view laser-scanning confocal micrographs of three particle stacks containing 23, 24, and 14 particles respectively (the third stack fractured upon impact of the 14th particle). The circles below indicate the size of the largest and smallest particle bonded within the deposit (original particle size). The fracture surface of the third deposit reveals three particles (colored), visualizing the lateral offset between particles. **b)** Corresponding line profiles along the vertical axis. The diameter of the deposits is 2–4 times the particle size.

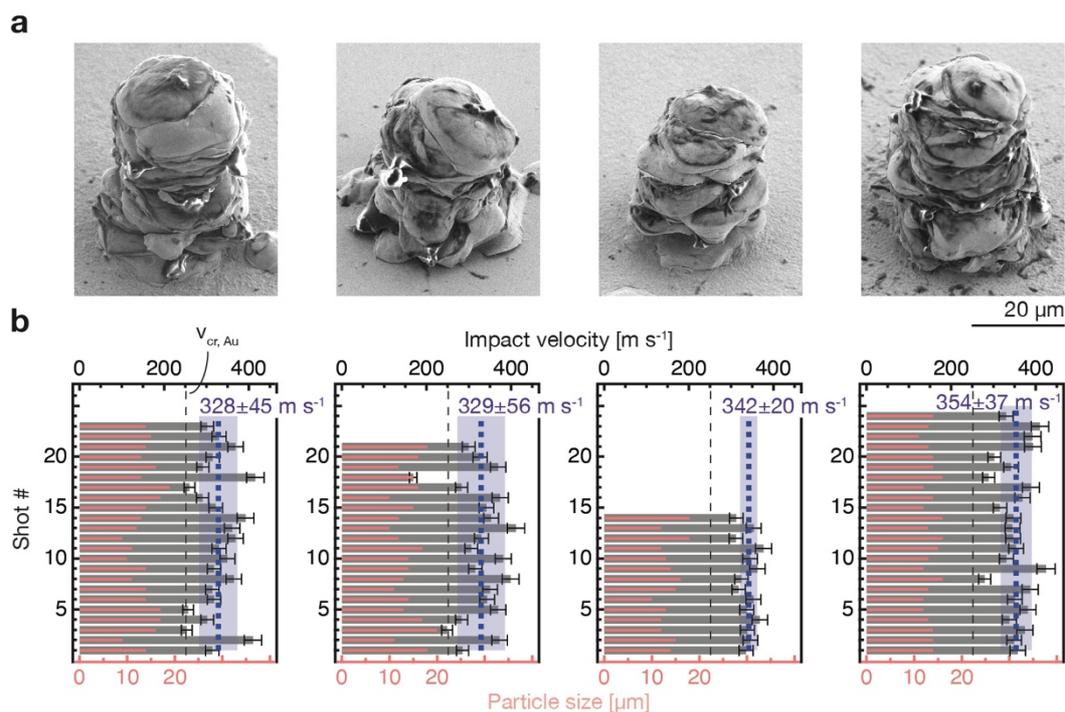

**Figure 3. Velocity and particle size spectra of deposits. a)** SE SEM micrographs of intact Au deposits (Viewing angle 45°, no tilt correction). **b)** Impact velocities (grey bars) and diameters (pink bars) of all particles shot at the respective target to build the corresponding deposit (note that some of the slowest particles might not have bonded, of which we do not have evidence). Error bars for velocity: ±5%. The dashed line labels the average velocity and the purple band the standard deviation. $v_{cr, Au}$=253±7 m s⁻¹ (black dashed line) denotes the critical velocity for bonding of single Au particles 16 µm in diameter[5].

# 4. Impact-induced erosion of deposits

Impact-induced erosion of cold-sprayed coatings during deposition should be minimized for maximum deposition efficiency. Different types of erosion events are typically associated with three velocity regimes[18–20,23]. Just below the critical velocity $v_{cr}$ for particle adhesion, the substrate may erode due to impacting but rebounding particles. At velocities larger than $v_{cr}$, particles start to adhere and a deposit is formed (with a deposition efficiency that increases with velocity). In this stage, previously adhered particles may be eroded due to impact-induced fracture (sometimes likened to grit blasting)[17]. At the upper end of the spray window, erosion of the coating and the substrate due to various high-energy-impact-induced phenomena is observed[19]. These high-velocity types of erosion—melting, hydrodynamic penetration and/or particle fracture—have been reported and quantified at the single-particle level in previous LIPIT studies[14,24,25]. In this report, we can now observe erosion from the second, lower velocity-regime using LIPIT. The growth sequence in **Figure 4a** documents particle-by-particle buildup until shot 10. Yet, upon shot 11, fracture of the previously deposited material is obvious. The post-fracture micrograph in **b)** indicates that inter-particle fracture occurred.

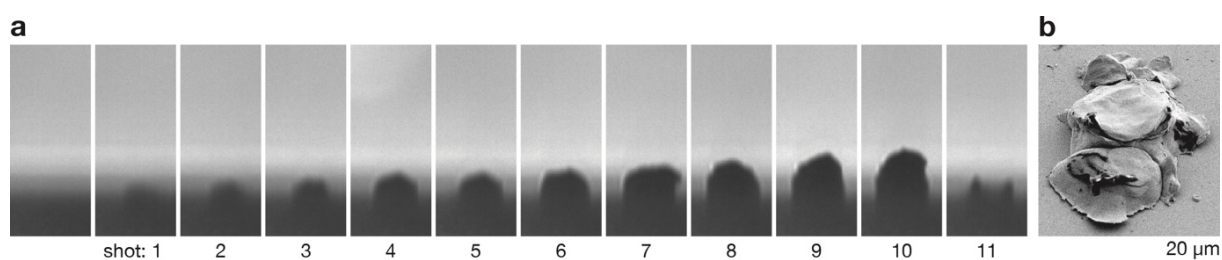

**Figure 4. Fracture of a deposit upon impact. a)** Sequence of optical micrographs that show the growth and fracture of a particle stack. **b)** Corresponding post-mortem SE SEM micrograph indicating inter-particle fracture (viewing angle 45°).

Fracture has been observed during the build-up of a number of deposits (**Figure 5a**). Sometimes, deposits fractured, continued to grow, and fractured again (the black arrows mark impacts that led to erosion as observed in the optical microscope). In almost all samples, the first fracture occurred before the 12th particle. At this stage, the deposits have an aspect ratio of roughly 0.5 or lower (as deduced by comparison to Figure 2, where deposits built from >20 particles have an aspect ratio of approximately one). Average particle velocities in fractured deposits—274±31 to 329±68 m s⁻¹ (purple band)—were consistently equal to or below the range of average velocities of successful coatings (green band, >328 m s⁻¹). From this data it is concluded that continuous buildup only occurred for an average particle velocity of ca. ≥330 m s⁻¹. Below this value, all of the current deposits fractured. Notably, an average particle velocity of >328 m s⁻¹ is well above the critical velocity $v_{cr}$=253±7 m s⁻¹ for bonding of individual particles (D=16 ± 4 μm)[5] (and so are the velocities of most individual particles in these coatings). This data thus suggests the existence of a minimum velocity for material buildup that is higher than the average critical velocity for bonding of single particles.

Multi-particle LIPIT experiments such as these may be particularly useful to quantify the contribution of erosion to the deposition efficiency at the lower end of the deposition window. In that range, researchers report that the deposition efficiency of the cold-spraying process is markedly below 100% after the onset of deposition, and only gradually increases with particle velocity within a transition zone that can span a few hundred meters per second[23]. Erosion due to blasting might contribute significantly to the slow increase in coating efficiency. But in cold-spray experiments, this contribution is challenging to decouple from the effect of particle-size-dependent critical velocities for adhesion[6] which—in combination with the ever-present size distribution within powders—naturally causes a gradual instead of a steep increase in deposition efficiency[23]. In addition, differently sized particles would be expected to have different velocities and temperatures as well, so the notion of a single 'critical velocity' at the particle level is very challenging to map to coating efficiency. LIPIT, by contrast, offers potentially straightforward decoupling of these effects (as particle size, velocity and temperatures can be separately controlled). In fact, our experiments suggest that erosion could lower the deposition efficiency by several dozen percent—Figure 4 and Figure 5 indicate that material loss by erosion could be more than 50%.

Despite the promise of the current technique, two points need to be taken into consideration when comparing the fracture events seen here to the particle blasting that can occur in cold-spray. First, and most importantly, these LIPIT deposits lack lateral constraints as they are built in isolation on a flat substrate. Deposits might thus experience impact-induced stress-states that are not observed in laterally-constrained cold-spray coatings (even though fracture occurs at low aspect ratios of ≤0.5). Future experiments might consider the addition of a random raster within a defined region in the plane to broaden the width of deposits and address this issue. Secondly, although we have no direct evidence to this effect here, there is a chance of organic contamination from the polymer launch pads used in the present LIPIT experiments, which may lower the bond quality. Deposition of organics atop of adhered particles and particle stacks is obvious in the SEM micrographs, and such layers may clearly have an undesirable effect for future bonding events, which are quite influenced by interface layers at the particle level[7,8,26].

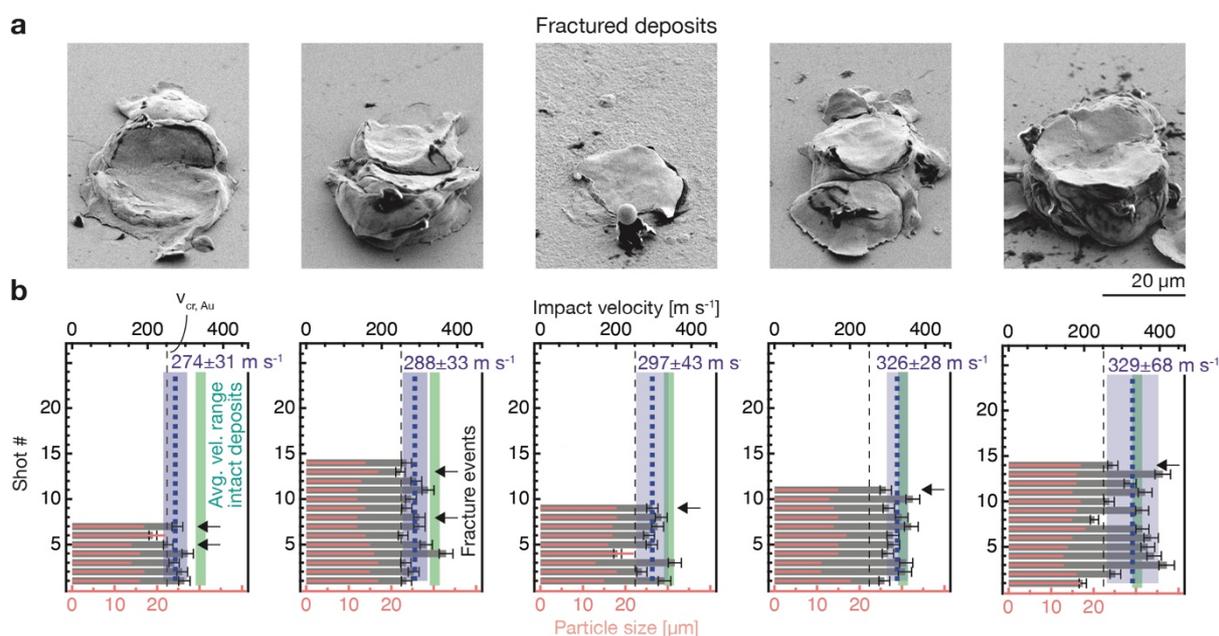

**Figure 5. Velocity spectra suggest a critical velocity for effective coating buildup. a)** SE SEM micrographs of fractured Au deposits (Viewing angle 45º, no tilt correction). **b)** Corresponding velocity and size spectra. The vertical green bar denotes the range of the minimum and maximum average velocities of the deposits from **Figure 3**, all of which were coherent deposits. The dashed purple line and band denote the average velocity and their standard deviation for each of the present experiments, which failed to produce a coherent deposit. Black arrows mark impacts that led to fracture as observed in the optical microscope. It can be noted that continuous buildup of the deposits is only observed for an average particle velocity of ca. ≥330 m s⁻¹. Significantly below this value, deposits fracture invariably. This suggest the existence of a critical velocity for material buildup that is higher than the average critical velocity for bonding of a single individual particle (the critical velocity for bonding of single Au particles, 16 μm in diameter[5], $v_{cr,Au}$=253±7 m s⁻¹ is marked by a dashed, black line).

## 5. Inter-particle porosity

Cross-sections of cohesive deposits—prepared by focused ion beam (FIB) milling—often reveal internal porosity (**Figure 6a**). As this porosity is characteristically found at particle-particle interfaces, we attribute these flaws to incomplete bonding. In principle, the degree of bonding might be correlated to the kinetic parameters of the particles involved. For example, one could make a first hypothesis that the pronounced gap in Figure 6a (black arrow) has been formed because of insufficient impact velocity of the group of particles located directly above the gap. In general, low velocity should reduce both bonding of the first particle, and the degree of peening that could promote closure of any porosity in the vicinity. Other related hypotheses could seek such a correlation with, e.g., both the size and velocity of the particles, or the kinetic energy of the particles. With the single-particle data provided by LIPIT, we proceed to develop an approach to quantitatively test such hypotheses.

First, the location of the pores throughout the volume as a function of deposit height is extracted by analysis of four cross-sections cut at approximately equidistant positions (at 1/5, 2/5, 3/5, and 4/5 of the deposit diameter) (Figure 6a). Specifically, we map the length of the internal flaws (measured by tracking the lower edge of each pore/fissure). Their projection onto a single plane is shown in **Figure 6b**. This spatial mapping of inter-particle flaws has been performed for three cohesive deposits (**Supplementary Figure 1**). The flaw length describes the extent of non-bonded particle interface and thus captures a variable that is likely affected by the impact of the individual particle[25,27]. It is thus more relevant in the present situation than, for example, the pore area. The pore area may be significantly affected by deformation of the stack during subsequent impacts (opening or further closing of the pore between non-bonded interfaces).

Second, we attempt to connect local porosity to impact parameters of associated particles. Because we are not able to map the precise location of individual particles from the LIPIT sequence in these stacks, we instead attempt to map *relative* positions in the stack to the particle sequence. Specifically, we plot the LIPIT particle characteristics (velocity or kinetic energy) as a function of accumulated total particle diameter of all the launched particles (**Figure 6c**). Here, the width of each horizontal bar corresponds to the original particle diameter and represents the relative contribution of each particle to the total height of the deposit. Then, this plot is scaled to the actual height of the deposit by aligning the bottom end with the lowest point of the bottom-most particle, and the bottom edge of the top data point with the approximate bottom edge of the topmost particle. We analyze the density of flaws based on this new height scale by dividing the pore map into slices that correspond to these relative particle diameters. Within each slice, the total flaw length is measured and normalized by the slice area (purple) (**Figure 6d**). With this display, we can connect density of coating flaws to a group of particles that have interacted with the relevant portion of the deposit, even though the exact particles involved in the production of the flaw cannot be pinpointed.

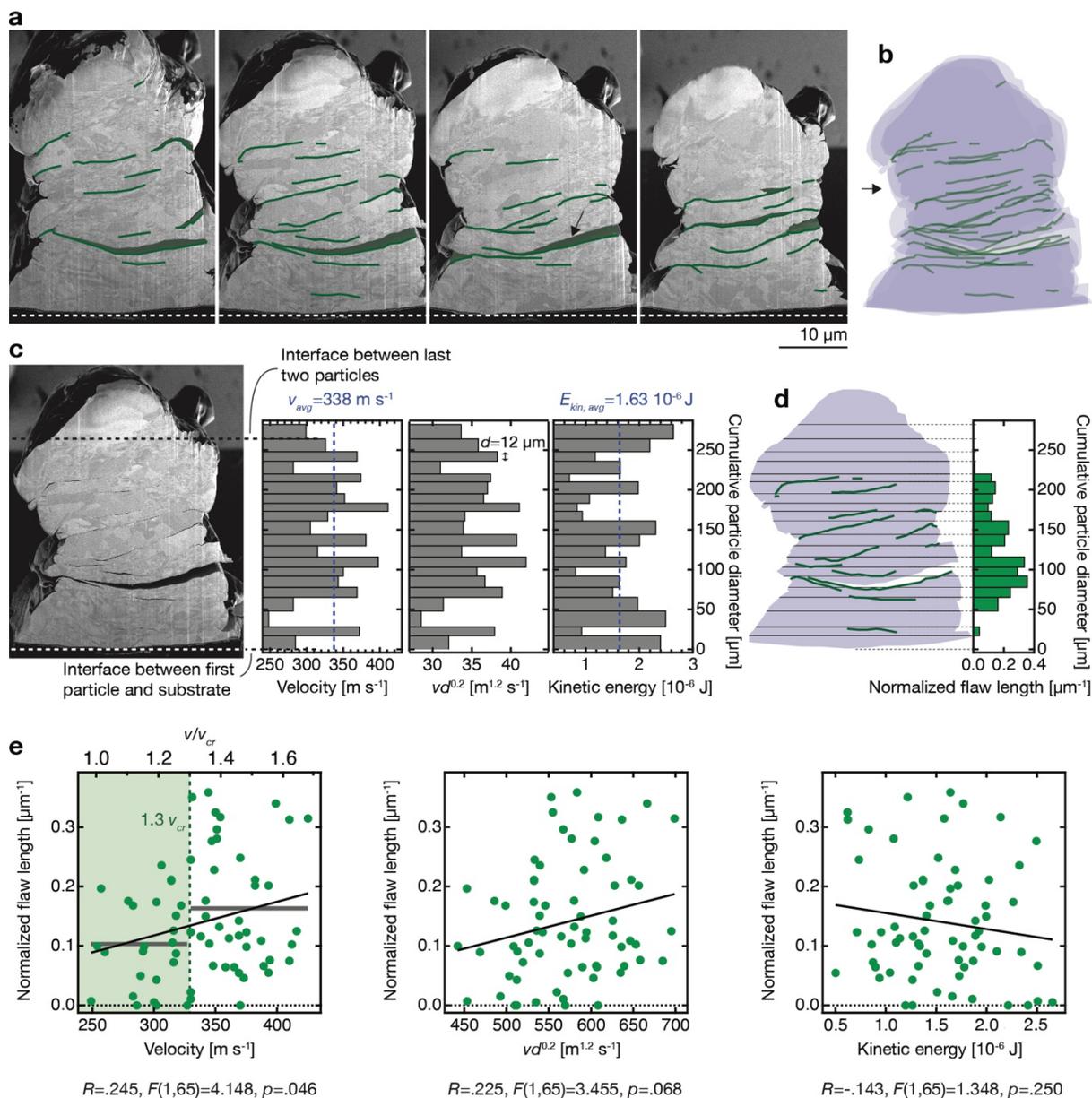

**Figure 6. Correlation between inter-particle defects and local particle impact characteristics. a)** SE SEM micrographs of FIB cross-sections cut at roughly equidistant intervals across the width of a deposit. Inter-particle porosity is highlighted in green, with the bottom pore interface marked with a dark-green line. The length of this interface is used to quantify the porosity. **b)** Overlay of the pore interfaces (green) extracted from the four cross-sections (outlines of the stack in purple). **c)** Velocity, velocity normalized by $d^{-0.2}$, and kinetic energy ($mv^2/2$) of individual particles plotted as a function of cumulative particle diameter—the width of each bar indicates the original diameter of individual particles before deposition. The plots are scaled to the total height of the deposit: the bottom end of the plot is aligned with the lowest point of the bottom-most particle, and the bottom edge of the top data point is aligned with the bottom edge (approximately average height) of the topmost particle. **d)** The density of flaws is quantified on the basis of the same height scale to enable direct comparison of the data to particle impact characteristics. The pore interface map is partitioned into slices with thicknesses that correspond to the relative particle diameter. Within these slices, the total length of flaws is summed for the respective portion of the stack and normalized by the slice area (only one out of four cross-sections shown). **e)** Scatter plots relating the normalized, total flaw length (combination of data from three stacks) to three different parameters that quantify individual particle impacts: velocity (and velocity normalized by critical velocity for bonding $v_{cr}$), velocity normalized by $d^{-0.2}$, and kinetic energy. The black solid line is a linear fit to the data. The horizontal grey lines in the first graph mark average values below and above 1.3 $v_{cr}$. A linear regression analysis indicates a significant correlation of increased flaw length with increased impact velocity ($R$=0.245, $F$(1,65)=4.148, $p$=0.046).

Thus, we can proceed to test the hypotheses laid out above, seeking correlations between areas of increased porosity with anomalies in the kinetic attributes of the relevant particles. The first hypothesis above would connect particle velocity with flaws, so particle absolute velocity is the first of the quantities plotted (Figure 6c). Yet, the critical velocity for particle adhesion is known to be particle-size dependent. Correspondingly, it should be assumed that the percentage of bonding and extent of peening may be a function of not only velocity but also particle size. Using LIPIT of Al and Ti, Dowding *et al.*[6] have experimentally verified an exponential scaling of the critical velocity $v_{cr}$ with particle diameter $d$, $v \propto d^{-n}$, with an approximate scaling factor of $n$=-0.2. Thus, we present the particle velocities normalized by $d^{-0.2}$ in a second plot. Finally, we also calculate the kinetic energy of impacting particles ($mv^2/2$).

With this approach, we can directly align kinetic properties with flaw location in each of our stacks, as illustrated in Supplementary Figure 1. Inspection of those data manually does not reveal correlations obvious to the eye, although a subtle correlation is indeed present and can be revealed by a linear correlation analysis of the combined data from all stacks (N=67). Such an analysis is shown in **Fig. 6e**, indicating that an increased impact velocity causes a weak (by 0.01 for every 10 m/s) but statistically significant ($\alpha$=0.05) increase of local porosity ($R$=0.245, $F(1,65)$=4.148, $p$=0.046). 6% of the variance in local pore length may be allocated to an increase of the impact velocity ($R^2$=0.06). No statistically relevant effects of $vd^{0.2}$ or the kinetic energy were found in the same analysis (note the p values above 0.05 in Fig. 6e).

Thus, this data suggests that a high impact velocity correlates with increased porosity. In contrast, a low particle-velocity by itself is not a factor in porosity development, even when that velocity is normalized by particle size in a manner expected to correlate to bonding characteristics. While this result is surprising at first, it does match well with expectations based on prior single particle impact experiments that probe bond development in Cu, another fcc metal. We can compare the present results to those for Cu by normalizing the velocities by the critical velocity for bonding $v/v_{cr}$, which is shown using the upper (secondary) x-axis in Fig. 6e. As this presentation reveals, we have measured impact velocities of 1–1.7 $v_{cr}$ ($v_{cr}$=253±7 ms$^{-1}$ for D=16 ± 4 μm[5]). This is a very broad range, which actually spans above the "optimum" velocity for single particle bonding, and crosses into a range where interparticle flaws increase due to the extreme deformation that results during the impact event. For example, evaluations of the metallurgically bonded area fraction of impacted Cu particles in both 2D[25] and 3D[27] report a maximum in bonding (i.e., a minimum total length of the interfacial pore in the present terminology) at a velocity of ~1.3 $v_{cr}$. Reported flaw density quickly increases to double the values at ~1.5 $v_{cr}$ and higher at >1.6 $v_{cr}$.

In light of these literature trends, the results in Fig. 6e make more sense: our work shows that in complex particle stacks, at 1–1.1 $v_{cr}$, the total pore length is lower than at 1.5 $v_{cr}$. Thus, the notion that flaw length increases at >1.3 $v_{cr}$ is supported by these results. We do not see an obvious increase and then decrease in bonding (or a minimum in flaw density at 1.3$v_{cr}$). On one hand, this may be due to sampling bias: as noted in the previous section, at lower velocities the incidence of fracture/erosion events is much higher. For flaw analysis such as in Fig. 6, we therefore preferentially sample stacks that survived longer, which in turn implies that we sample higher velocities preferentially to lower ones. On the other hand, we deal with porosity that is likely not only a result of the immediate impact but is also affected by down-stream peening. Thus, at least part of the statistical scatter should originate from variations not only of the immediate impact velocity but also in down-stream peening intensity.

Our results on fracture and flaws, taken together, suggest that the window for optimal deposit development may be extremely narrow. Whereas single particles bond at $v_{cr}$ and optimize their bonding at 1.3$v_{cr}$, the particle deposits in this work require higher velocities to avoid fracture, already close to 1.3$v_{cr}$ (~328 m s$^{-1}$), above which bonding levels drop and flaw densities rise.

This result suggests that work which can narrow the range of particle velocities would be especially relevant for the production of low flaw-density coatings by cold spray.

Finally, the scatter of the data indicates that velocity can only be one factor in bonding or densification. Likely, other local conditions, such as precise offsets of impacts, specific stacking sequences, or specific deformation behavior of particles, need to be considered. For example, it seems possible that a pronounced, local protrusion—for example the inclusion of a below-average-size, satellite particle, or the upward extrusion of material—in combination with a local dip in kinetic energies may have favored the development of the extensive gap in a stack shown in **Supplementary Figure 2**.

## 6. Microstructure evolution

Cold spray coatings typically feature heterogeneous microstructures that are characterized by extreme strains, strain rates and strain gradients. Deformation, work-hardening and the resulting recovery and recrystallisation—both dynamic and static (during deposition or a post-deposition heat treatment)—are the processes that underlie the formation of a typically complex, bimodal grain structure. Based on the analysis of full coatings as well as individual particle splats, the heterogeneous microstructure is generally attributed to a strongly heterogeneous distribution of strain within deposited particles[2,28]. FEM simulations[29] and experimental data[15] estimate the strain in peripheral regions of particles to be up to 5 times higher than in the center. While the importance of the intra-particle strain gradient cannot be underestimated, one needs to assume that there are also significant differences in degree of deformation between individual particles that originate from size- and spray-dependent variations in impact velocity. In general, the size-dependence of particle acceleration can cause a variability of impact velocity by tens of percent (and by extension, a variability in strain) within a particle batch of typical size distribution[18,30]. Further, pronounced differences in particle deformation for impacts at the center and near the rim of the spray have been demonstrated[31]. However, establishing the effect of the significant velocity distribution on the final microstructure is difficult at the coating level, an issue which can be addressed with the present approach.

**Figure 7** presents three characteristic microstructures that are found within all particle stacks: first, volumes of heavily deformed, micrometer-scale grains that represent the original but work-hardened microstructure of the particles (red. The large lattice strains show as distorted, gradient grayscale contrast in the BSE SEM micrographs); second, volumes of large, stress-free grains that must be the result of extensive grain growth upon recrystallization (blue); and finally a seemingly transitional regime (orange) that contains, on one hand, what appear to be well-formed recrystallization nuclei (areas with typically sub-micrometer grain size and low to medium distortion (black arrow in b)), and on the other hand, large regions of heavy deformation of the prior microstructure (red arrows).

These three microstructural types correspond rather well to the conventional individual stages of strain and recrystallization seen in other mechanical working processes. The heavily strained, large crystallites (red) that most likely are residual from the original particles correspond to "stage 1": accumulation of plastic deformation in the prior micron-scale grains. At the other end of the process, fully recrystallized, stress-free volumes (blue) correspond to a conventional "stage 3". And the areas that show signs of stress-relaxation by recovery and nucleation of new, submicron-scale grains (orange) likely belong to an intermediate "stage 2". One open question is the duration and timing of the transitions between these stages. In contrast to the cold-spray process, in LIPIT there are minutes between individual impacts, and the temperature of the deposit can be assumed to be room temperature for all but the shortest times (no post-deposition heat treatment was performed). Consequently, no particular time or temperature can be pinpointed as the trigger for recovery and recrystallization. Related to the question of timing is the question of whether there are multiple cycles of the sequence—whether stage 3 material is reworked by subsequent impacts to produce a microstructure of stage 1 that then can undergo the same cycle again. This of course would require recrystallization in-between impacts, not after, and this may be less likely in room-temperature LIPIT experiments than in cold spray.

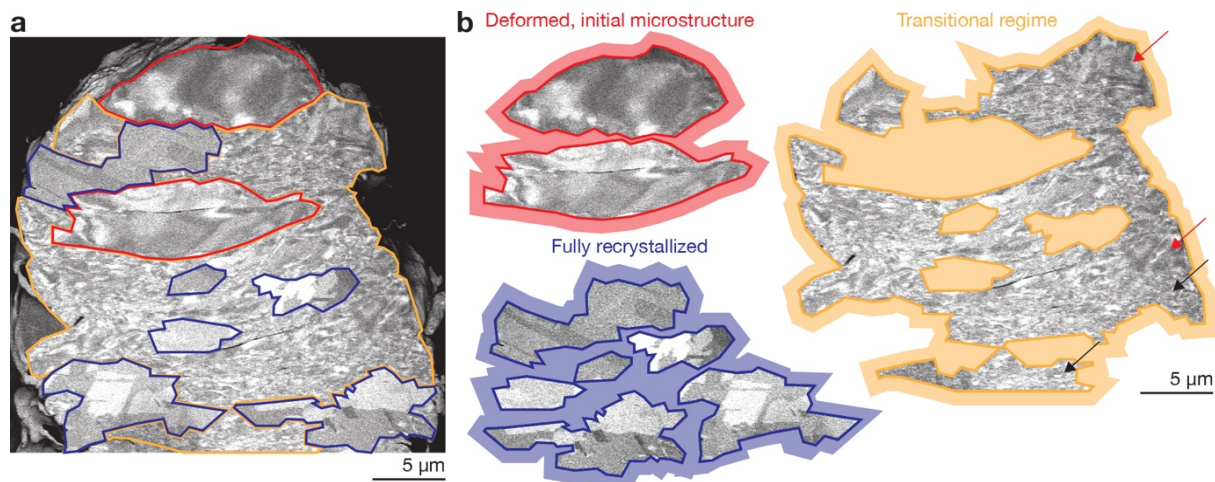

**Figure 7. Recrystallization and grain growth. a)** BSE SEM micrograph of a FIB-milled cross-section of a deposit (tilt corrected). Three distinct microstructures—found in all deposits—are highlighted in a) and isolated in **b)** Red: areas of large grain size with high defect density and lattice strains (characterized by distorted, gradient grayscale contrast)—the original but heavily deformed microstructure of the particles. Blue: fully recrystallized, large, stress-free grains. Orange: intermediate stage that contains what appear to be well-formed recrystallization nuclei (black arrows), as well as large regions of heavy deformation of the prior microstructure (red arrows).

An analysis of the volumetric distribution of these three microstructures is made in two of the four cohesive particle stacks (**Figure 8, Supplementary Figure 3**); the two deposits featuring the most prominent gaps were excluded. Several interesting observations are apparent from these views. First, the initial but heavily strained microstructure is most prevalent in the top-most particle. This is an important point when thinking about single-particle studies of microstructure evolution in cold spray[15,32,33]. Although a single particle impact produces considerable deformation with attendant interesting details, a substantial portion of its microstructural evolution history is certainly a product of the subsequent peening it receives. In fact, it may be difficult to connect single particle observations of, *e.g,* recrystallization or other structural change with the microstructures seen in cold spray generally in light of this observation.

Second, it is curious to note that some of the deformed prior-particle microstructure (red) survives all the subsequent deformation imposed from downstream impacts, and remains in lower portions of the stacks. The red regions in Figure 8 or Supplementary Figure 3 are often buried rather deeply. At the same time, recrystallized areas (blue) are sometimes located above recovered (orange) volumes. This suggests that the microstructure evolution is not merely a function of the number of impacts that have peened a particular volume. Assuming that, to first order, the degree of recrystallization is representative of the degree of deformation and cold work, it follows that there is no simple gradient in degree of cold work from the top to the bottom of the stack, nor is there an obvious "steady-state" deformation condition developed over the duration of the present experiments. If there were a deformation gradient, we would expect to find the sequence *red-orange-blue* with blue at the bottom (peened the most) and red at the top (not peened at all)—a sequence we do not find. If a steady state had developed the microstructure would be uniform at regions below the near surface. Instead, red and blue portions are not distributed randomly and uniformly throughout the stack but tend to group with their own at certain heights. The process of particle stacking thus clearly involves a stochastic process of freezing local strain concentrations based on nuanced local variations in particle placement, microstructure, and—perhaps—particle kinetic properties. We turn our attention to this last point in what follows below.

The local variations in microstructure are quantified as modulations of the respective area fractions as a function of stack height (**Figure 8c**). **Supplementary Figure 4** describes the analysis in detail, which follows the same approach already used to quantify local porosity. The local area fractions can then be compared to the local particle kinetics in **Figure 8d**. Both datasets suggest a spatial correlation between microstructural features and the impact velocity of involved particles. Non-recrystallized volumes (red) seem to be co-located with groups of particles of (mostly) below-average velocity, and fully recrystallized grains (blue) with a number of high-velocity impacts. In particular, all impacts of velocity significantly above average (ca. >370 m s$^{-1}$) can be connected with recrystallized areas. These qualitative observations are substantiated by a linear regression analysis of the combined data of both stacks (N=47) **(Figure 8e)**. An increase in impact velocity causes a statistically significant increase in recrystallized (blue) area fraction ($F(1,45)=4.472$, $p=0.040$) (by 0.01 for every 10 m/s), and a decrease in the area fraction of the initial microstructure (red) ($F(1,45)=4.316$, $p=0.043$) (also by 0.01 for every 10 m/s). 9% of the variance in microstructure area fraction is explained by the particle velocity (blue: $R^2=0.090$. red: $R^2=0.088$). No significant effect of velocity on the orange area fraction was discerned. Similarly, no significant effect of kinetic energy or $vd^{0.2}$ on either area fraction was found ($p>0.05$) (**Supplementary Figure 5**).

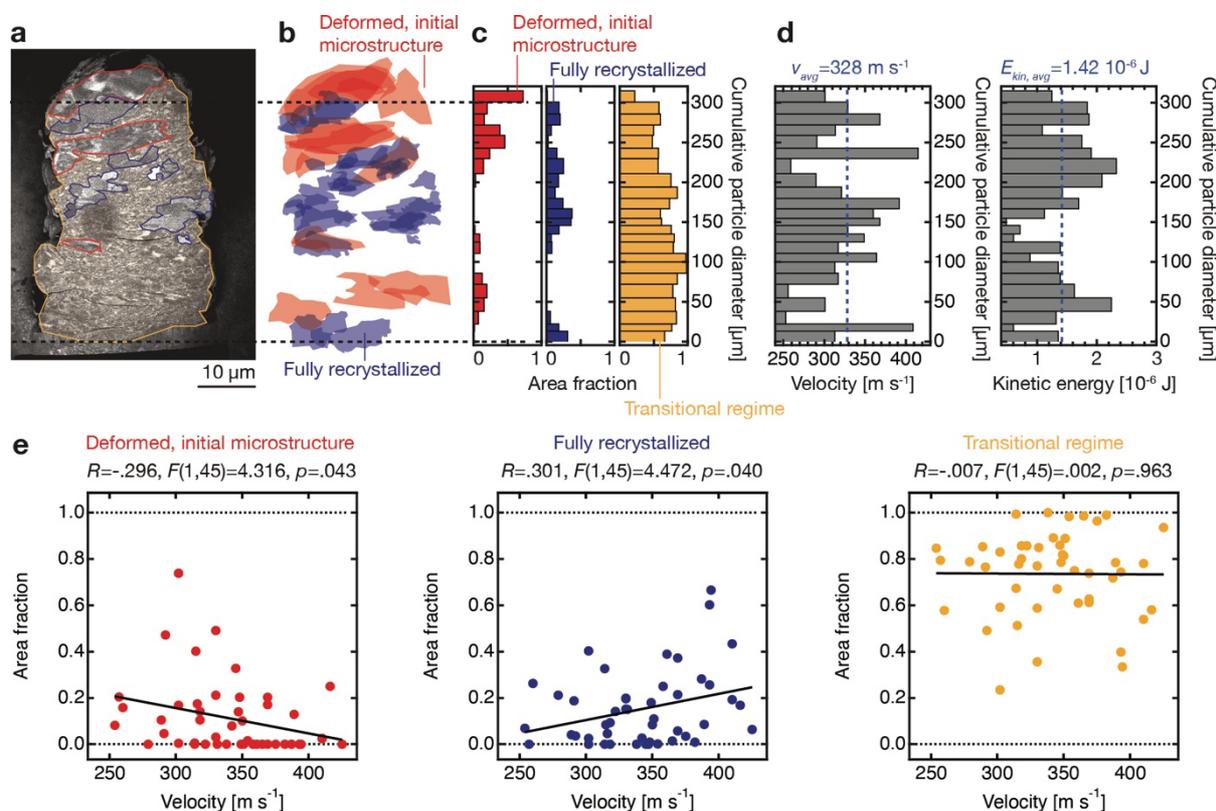

**Figure 8. Correlation between microstructures and local impact velocity. a)** Cross-sections (BSE SEM, tilt-corrected) of a particle stack with characteristic microstructures outlined. **b)** Microstructure maps assembled from analysis of four cross-sections per stack. Red areas represent the heavily deformed but original particle microstructure. Blue are areas of stress-free,

micrometer-scale grains (the recrystallized microstructure that underwent grain growth). For clarity, orange sections were omitted in these presentations. **c)** Area fractions for each microstructure group as a function of stack height (cumulative particle diameter). **d)** Spectra of velocity and kinetic energy of the individual particles plotted as a function of cumulative particle diameter. The plots are scaled and aligned to the total height of the deposit, as in Figure 6. The dataset hints towards a spatial correlation between non-recrystallized volumes (red) with groups of particles of (mostly) below-average velocity, and stress-free, microscale grains (blue) with a number of high-velocity impacts. **e)** Scatter plots relating the area fractions of the three different microstructures (combined data of two stacks) to the local impact velocity. The black line is a linear fit to the data. The effect of particle velocity is negative on the fraction of initial, deformed microstructure (red) ($F_{(1,45)}$=4.316, $p$=0.043); but is positive on recrystallized area fraction ($F_{(1,45)}$=4.472, $p$=0.040). No correlation between local impact velocity and the intermediate microstructure (orange) is found.

That faster impacts lead to more evolution towards recrystallization makes sense intuitively in two complementary ways, both of which are about strain. Faster impacts should produce more strain, which increases (i) the amount of stored defect energy, providing regions that are more poised to recrystallize, and (ii) local adiabatic temperature rise, which goes with strain and promotes faster recrystallization kinetics. Since these phenomena both connect directly to strain, it is useful to attempt an analysis purely based on microstructural data that relates the microstructure evolution to the degree of local deformation.

In **Figure 9a**, the local particle flattening is used as an estimate of local strain as is common in the cold spray literature[12], by comparing splat heights along the centerline of two stacks (**Supplementary Figure 6**). We only measure splat heights instead of the ratio of splat height to width, because detecting the complete outlines of splats was not always possible; especially in areas of recrystallization, the interfaces were obliterated. Because a reasonable agreement between the flattening estimated by splat height and that estimated by splat aspect ratio was found (**Supplementary Figure 7**), we have used the less precise but more reliable strain estimation only based on splat height. Here, the vertical distance between particle-interfaces is mapped along the stack height (Figure **9**a) and averaged across four cross-sections to derive an average splat height $h$ versus stack height (**Figure 9b**). Using the average, original particle diameter of the particles in the stack, $D_{avg}$, $h$ is then converted to a nominal flattening parameter $\varepsilon_n$ that is akin to the nominal, average compressive strain along the deposition direction:

$$\varepsilon_n = \frac{D_{avg} - h}{D_{avg}}$$

For convenient comparison to microstructural area fractions, $\varepsilon_n$ is mapped to the same y-axis "cumulative particle diameter" (Figure 9b).

The area with highest strain is located at approximately half of the height of the stack (red shading in Figure **9**b). Notably, this region of highest strain correlates with the region of most pronounced recrystallization (blue in **Figure 9c**). In fact, a linear regression analysis of the dataset that combines two stacks (N=47) indicates that a higher strain causes an increase in total recrystallized area fraction (sum of fully recrystallized fraction and fraction of transitional regime) ($F_{(1,45)}$=12.016, $p$=0.001) (**Figure 9d**). 21% of the variance in recrystallized microstructure area fraction is explained by the nominal, local particle flattening parameter, that is, the local strain ($R^2$=0.211). It can be noted that the same analysis performed for only the fully recrystallized area fraction (blue) does not unveil a statistically relevant connection to strain. One reason for this might be the aforementioned difficulty to identify particle boundaries in fully recrystallized areas, which adds uncertainty to the correlation analysis of these areas with local strain values. Thus, a more precise method to identify particle outlines and thus infer local flattening ratios and strains could improve the dataset.

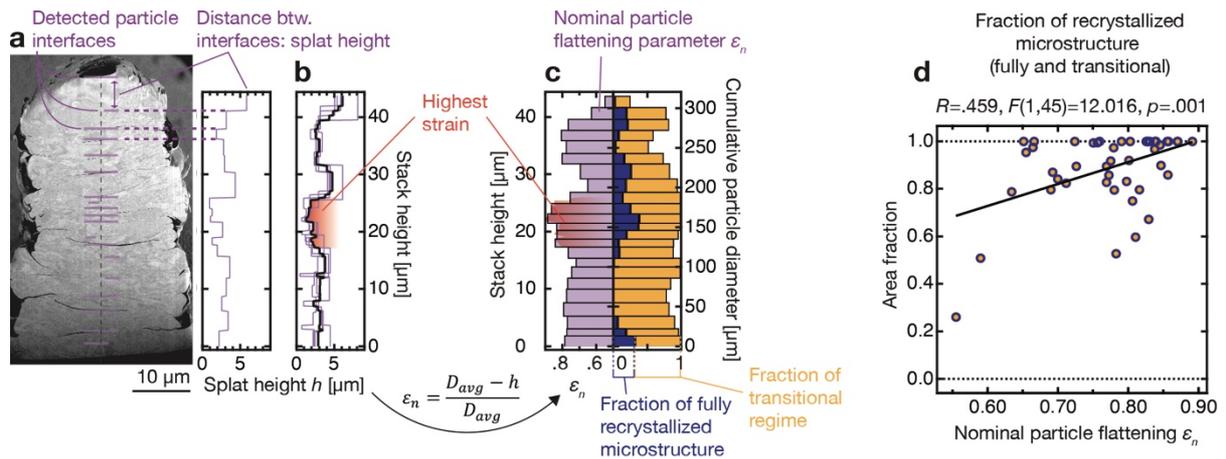

**Figure 9. Strain-dependence of microstructure evolution. a)** Cross-sectional micrograph (SE SEM, tilt-corrected) of a stack. Purple lines mark detected particle interfaces along the centerline of the deposit (off the center line if not otherwise possible). The graph plots splat heights $h$ (interface distance) as a function of stack height as measured in the single cross-section. **b)** The average splat height (black) for the whole volume, derived from four measurements in roughly equidistant cross-sections (purple). **c)** A nominal flattening parameter $\varepsilon_n$, akin to the nominal, average compressive strain along the vertical direction, is derived from $h$ and the average particle diameter $D_{avg}$. Mapping of $\varepsilon_n$ onto the y-axis "cumulative particle diameter" (purple bars) enables the direct comparison of local strain to the local area fraction of the two recrystallized microstructure groups "recrystallized microstructure" (blue) and "transitional regime" (orange). **d)** A linear correlation analysis (data from two stacks) demonstrates a strong effect of high strain on the prevalence of total recrystallized microstructure ($F_{(1,45)}=12.016$, $p=0.001$).

In combination, these results suggest that a local decrease in impact velocity (ca. 10–20%) can result in areas of decreased strain insufficient to trigger recrystallization. Correspondingly, a local increase of impact velocity (also ca. 10–20%) may result in areas that develop an accumulation of recrystallized volume. These results support the thought that at least part of the microstructural heterogeneity in cold-spray coatings may be caused by strain differences amongst impacted particles, and not only by strain gradients within particles. This points to an interesting direction for future work.

# 7. Conclusions

Here we have presented a many-particle LIPIT method that reproducibly stacks particles (>20 particles) with different characteristic spectra of impact velocity. The transition of LIPIT from single-particle testing to multiple-particle interaction problems also introduces a new element of stochasticity into LIPIT research generally. With the variability of particles, velocities, and impact surfaces that result, the results provide a bridge between ideally-characterized and nearly deterministic results at the single-particle level, with the uncharacterized multitudes of particles in a true manufacturing environment. Our results here show that statistical connections can be made between particle properties and the characteristics of the deposit, across many particles. We have thus validated a many-particle testing mode that we believe has great potential for the future study of a wide range of many-particle problems. The most salient results of this student include the following:

- We have observed the successful build-up of stacks but also their impact-induced fracture, which occurs preferentially at lower sequence velocities where the bonding is poorer and impacts thus cause breakage. Based on average velocity data, we concluded the existence of a velocity for successful accumulation of a coating that is higher than the critical velocity for adhesion of single particles (ca. 1.3 $v_{cr}$).

- We have found that individual particle velocity data can in some cases be statistically correlated with internal features of the coating. For example, local interfacial porosity (non-bonded particle interfaces) is correlated with particle velocity. This positive correlation may be a result of increased sampling of the "hydrodynamic penetration" regime of impact, which provides peeling forces that lower the adhesion of individual particles. Combined with prior single-particle data, one may set $v \leq 1.3\ v_{cr}$ as a range for optimal densification. Taken together with the minimal average velocity for coating buildup (>1.3 $v_{cr}$), this makes for a very narrow velocity band for optimal deposit development in these experiments, and may suggest a fruitful direction for cold spray optimization through narrowing of the statistical distributions of particle sizes and velocities in a spray.

- A statistically significant correlation between particle impact velocity, local strain, and recrystallized fraction hints towards an influence of strain-gradients between particles—and not only within single particles—on the heterogeneous microstructure of cold-spray coatings, where faster particles cause more strain and thus an increased local driving force for recrystallization.

These first-of-their-kind analyses should be directly applicable to similar studies in different materials or with larger numbers of particles. At the same time, the work motivates future experimental capabilities that address some of the shortcomings of this first demonstration. Rastering instead of mere stacking of particles should enable the production of deposit that better represent the low-aspect coating geometry. New analyses are needed to relate kinetic data of individual particles to exact particle positions inside the stack rather than just their proportional position. We look forward to a new era of multi-particle sequence experiments that explore the complexities of cold spray and other impact-related processes.

## Acknowledgments and funding information

**Acknowledgments:** The authors thank Steven Kooi, MIT, for designing and constructing the
current LIPIT setup and support with the system. Electron microscopy was performed at
MIT.nano.

**Funding:** Work on developing new LIPIT methodology and analysis is supported by the
Army Research Office and was accomplished under Cooperative Agreement Number
W911NF-22-2-0109; Work exploring microstructure evolution under extreme conditions is
supported by the U.S. Department of Energy, Office of Science, Office of Basic Energy
Sciences, Division of Materials Sciences and Engineering [DE-SC0018091]; A. Reiser was
supported by the Swiss National Science Foundation [Early Postdoc.Mobility Fellowship,
P2EZP2 188070].

## Author contributions

**A. Reiser:** Conceptualization; Methodology; Analysis and interpretation; Investigation;
Writing–Original draft; Visualization; Funding acquisition. **C. A. Schuh:** Writing–Reviewing
and Editing; Analysis and interpretation; Supervision; Funding acquisition.

## Graphical abstract:

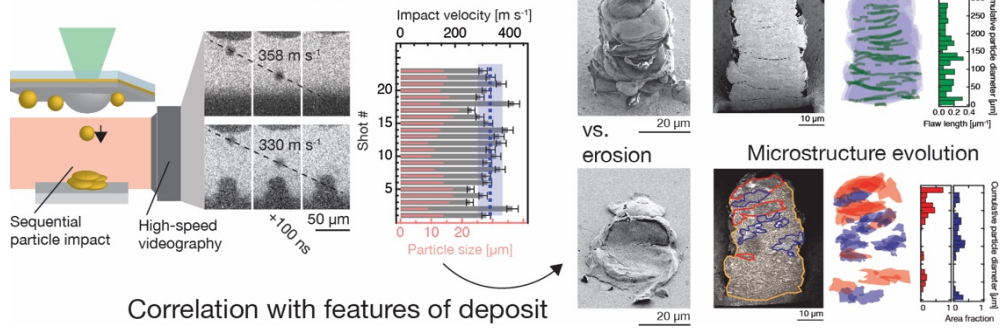

# 8. Supplementary Information

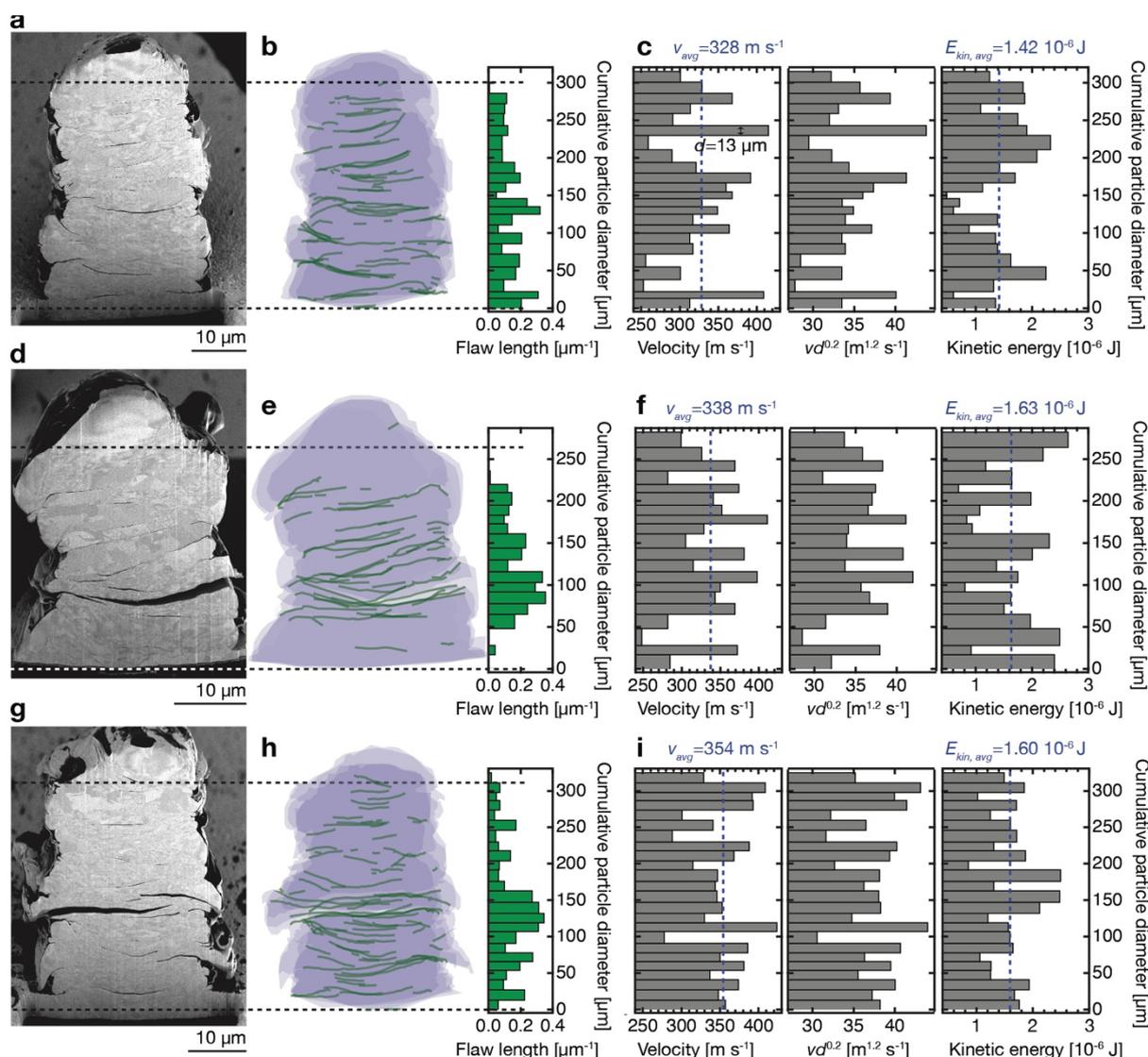

**Supplementary Figure 1. Comparison of the location of inter-particle defects to particle impact characteristics in individual stacks. a, d, g)** Representative SE SEM cross-sectional micrographs of cohesive deposits (tilt corrected). **b, e, h)** Flaw interface maps (overlay from four, roughly equidistant cross-sections) and total normalized flaw length as a function of cumulative particle diameter. **c, f, i)** Velocity, velocity normalized by $d^{0.2}$, and kinetic energy ($\mathrm{m}v^2/2$) of all individual particles plotted as a function of cumulative particle diameter. The average velocity and kinetic energy are indicated by a dashed, blue line.

A statistical analysis has suggested a connection of higher degrees of porosity with higher impact velocities. Given the possible error in spatial mapping of pores to particle velocities (a shift of LIPIT datapoints a particle diameter up or down should likely be considered as a possible error), it is worthwhile to argue against the opposite theory (that a lower impact velocity causes increased porosity) on an individual basis. In general, pronounced dips in velocity (or $vd^{0.2}$) do not obviously co-locate with notably higher degrees of porosity (Supplementary Figure 1). Along the same lines, the concentration of wide horizontal gaps (d, g) does not align with particularly low impact velocities. In stack (d), the first, second and fourth particle are of low velocity, but they are not likely spatially correlated with the porosity: in the corresponding cross-sections, at least three particles can be counted below the gap. As this is likely an undercount, the low-velocity particles have likely formed the portion below the gap. In stack g), low-velocity particle number eight is within the vicinity of the accumulation of flaws, but multiple particles must have formed the associated, porous volume (and particles seven and nine are of above-average velocity).

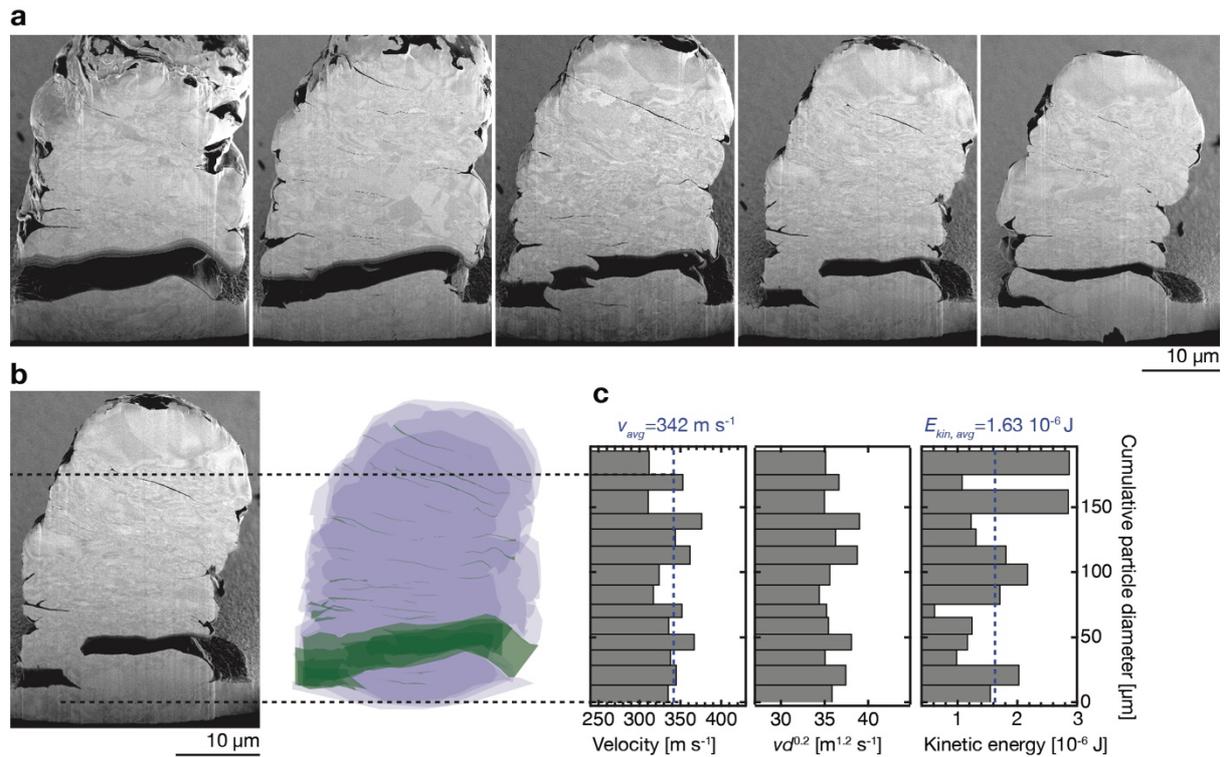

**Supplementary Figure 2. Pore formation. a)** A sequence of cross-sections throughout a cohesive particle stack and **b)** mapping of pores documents connection between upper and lower stack volume only through a small portion of material (fourth micrograph in (a)). It can be well imagined that such a protrusion—a small particle, or the upward extrusion of material from the particle impact below—that is impacted with larger, low-energy particles (as indicated by the energy spectrum in **(c)**) can result in a large gap as observed. Impacting particles might have sufficient energy to bond to the point of first impact (the protrusion) but not sufficient extra energy to also bond to the larger particles below—they rebound but remain attached to the protrusion that acts as a hinge. It should be noted that no small particle <10 μm was registered in the LIPIT experiments—it might have been a satellite particle connected to a larger particle.

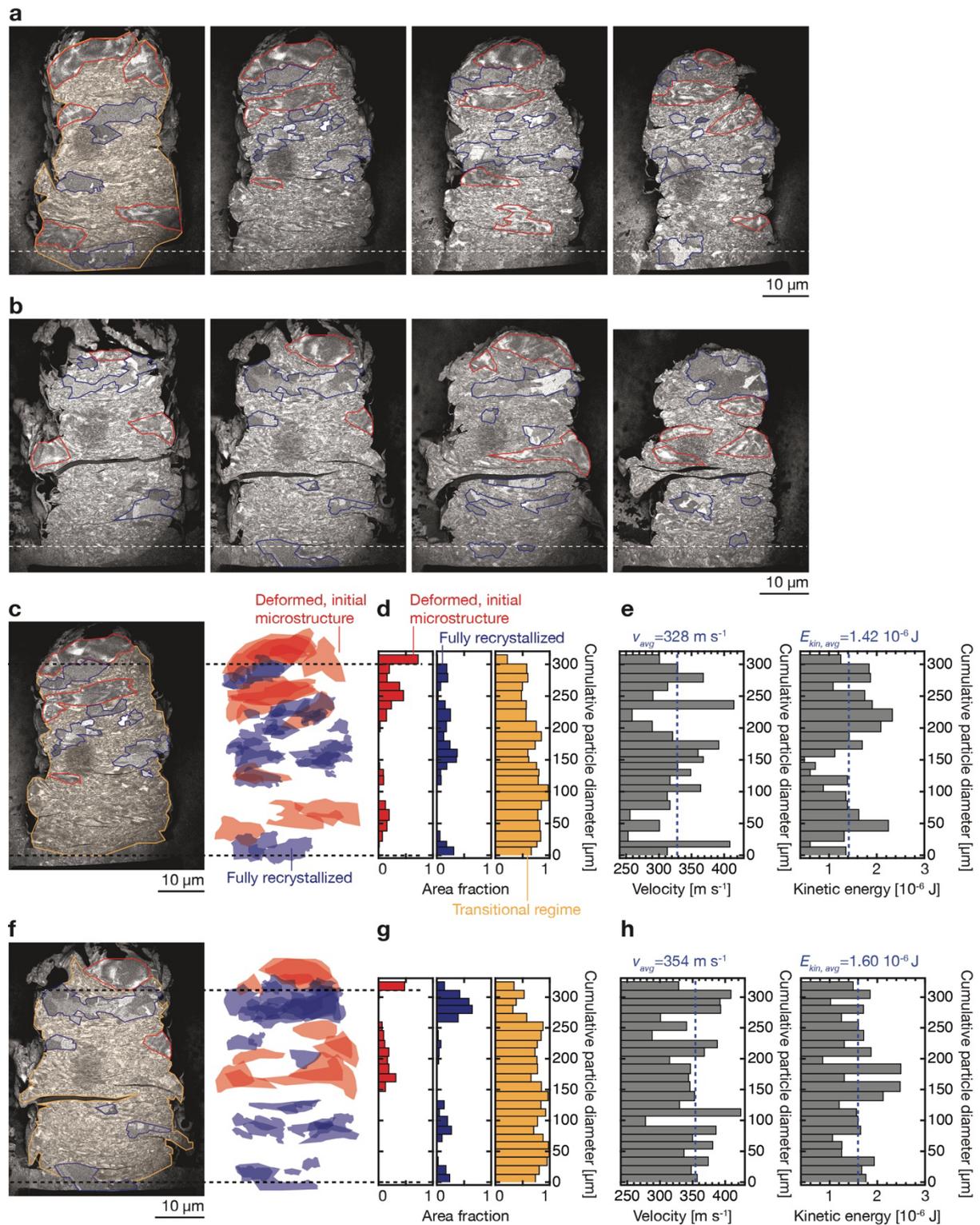

**Supplementary Figure 3. Microstructure of stacks. a, b)** The complete set of cross-sections (BSE SEM, tilt-corrected) of the two particle stacks shown in the main text, with characteristic microstructures outlined. **c, f)** Microstructure maps assembled from the analysis of the four cross-sections. Red areas represent the heavily deformed but original particle microstructure. Blue are areas of stress-free, micrometer-scale grains (the recrystallized microstructure that underwent grain growth). For simplicity, the transitional microstructure is only indicated in the first micrograph (Orange). **d, g)** Area fractions for each microstructure group as a function of stack height (cumulative particle diameter). **e, h)** Spectra of velocity and kinetic energy of the individual particles plotted as a function of cumulative particle diameter.

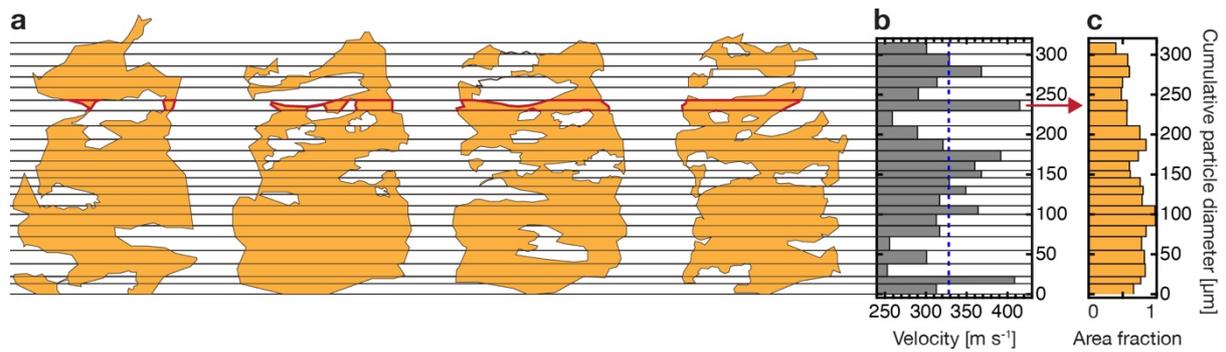

**Supplementary Figure 4. Calculation of area fraction as a function of cumulative particle diameter. a)** To quantify the
relative portion of the three characteristic microstructures as a function of stack height, the microstructure maps were first
divided into horizontal slices that equal the relative particle diameters from the LIPIT velocity spectrum (b) (with the spectrum
scaled to the stack height as described in Figure 6). The areas of the slices were summed to render a total area for each group
(orange, blue, red) at any given stack height (the area above the topmost, black line was assigned to the last particle). The
values were then normalized by the total area of all slices at respective height to give an area fraction (c).

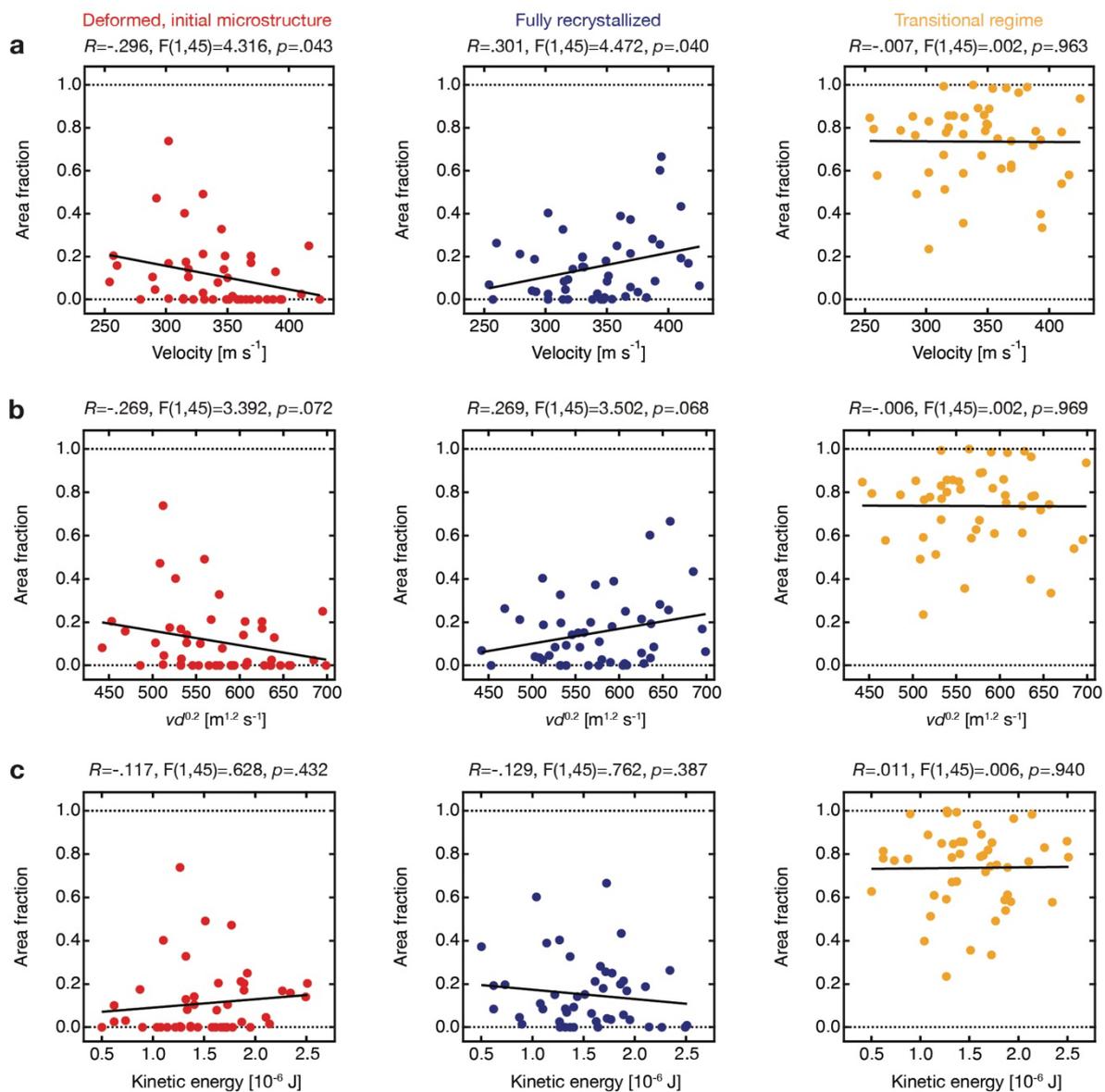

**Supplementary Figure 5. Linear regression analysis of local microstructure fraction and different particle impact
characteristics.** Scatter plots plotting the area fraction of the three microstructure types (combination of data from two stacks)
versus all three parameters that quantify individual particle impacts: velocity, velocity normalized by $d^{0.2}$, and kinetic energy.
The black solid line is a linear fit to the data. The results of a linear regression analysis are stated above the graphs. Only
impact velocity has a significant effect on the local microstructure ($p<0.05$).

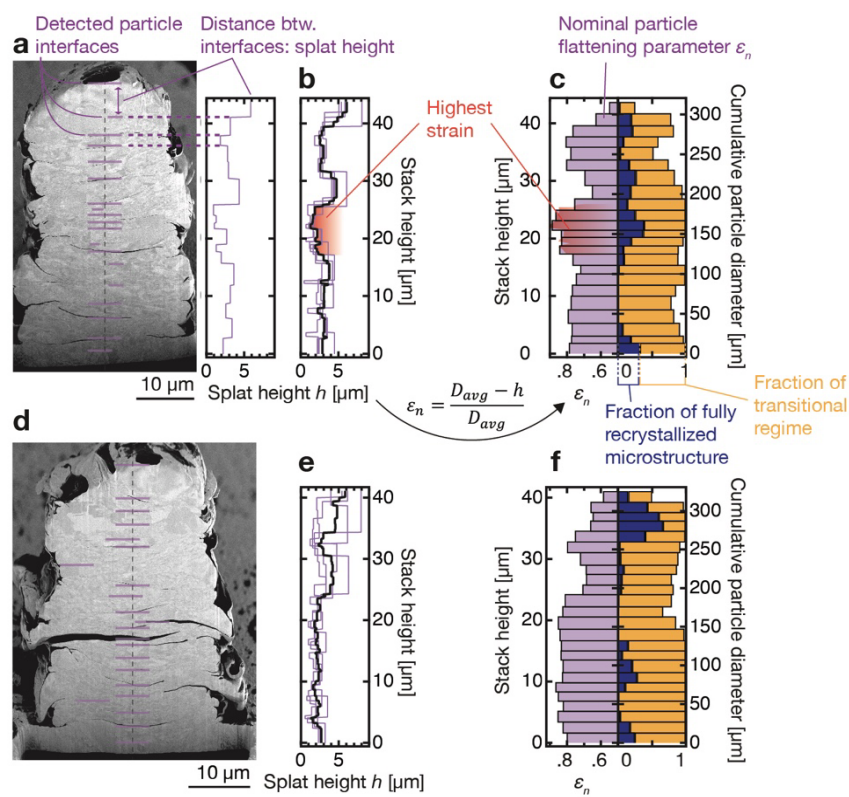

**Supplementary Figure 6. Strain-dependence of microstructure evolution in two stacks. a, d)** Cross-sectional micrographs (SE SEM, tilt-corrected) of two stacks. Purple lines mark detected particle interfaces along the centerline of the deposit (off the center line if not otherwise possible). The graph to the right of the micrograph in a) plots splat heights $h$ (interface distance) as a function of stack height as measured in the single cross-section. **b, e)** The average splat height (black) for the whole volume, derived from four measurements in roughly equidistant cross-sections (purple). **c, f)** A nominal flattening parameter $\varepsilon_n$, akin to the nominal, average compressive strain along the vertical direction, is derived from $h$ and the average particle diameter $D_{avg}$. Mapping of $\varepsilon_n$ onto the y-axis "cumulative particle diameter" (purple bars) enables the direct comparison of local strain to the local area fraction of the two recrystallized microstructure groups "recrystallized microstructure" (blue) and "transitional regime" (orange).

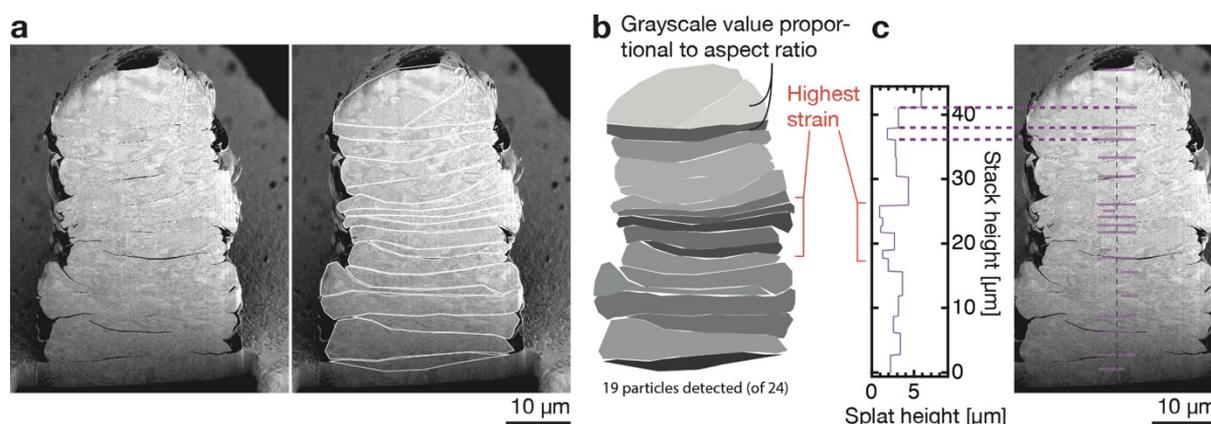

**Supplementary Figure 7. Comparison of local strain measurements. a)** Cross-sectional SE SEM micrograph with particle outlines highlighted in the right image. Full particle interfaces cannot always be seen. Here, approximate positions for the interfaces were assumed. **b)** Splats from (a) shaded in a greyscale value proportional to the aspect ratio of their bounding box (that is, flattening ratio, proportional to the strain). **c)** Splat heights from the same cross section, measured as described in the main text. Qualitatively, the two analyses are in agreement: highest strain is located in a volume at half height of the stack. Because of the general uncertainty in locating splat outlines as in (a), we have used the less precise but more reliable strain estimation as in (c).

23